\documentclass[onecolumn, amsmath, amssymb, aps]{revtex4-1}
\usepackage{xcolor}
\usepackage{graphicx}
\usepackage{dcolumn}
\usepackage{bm}
\usepackage[colorlinks=true,linkcolor=black,citecolor=red,urlcolor=blue]{hyperref}
\usepackage[mathlines]{lineno}
\usepackage{epsfig,lineno}
\usepackage{siunitx}
\usepackage{xcolor}

\begin{document}
\fontsize{14}{15}\selectfont
 
\title{POSSIBILITY TO STUDY PHOTON SCATTERING 
BY PROTON IN THE REACTION \Large $ep-ep\gamma$}

\author{A.I.~Lvov, V.A.~Petrunkin}
\affiliation{P.N. Lebedev Physical Institute Leninsky Prospect 53, Moscow 117924, USSR} 
\author{S.G.~Popov and B.B. Wojtsekhowski}
\affiliation{ Institute of Nuclear Physics, 690090, Novosibirsk, USSR}

\begin{abstract}
\large
 A theoretical possibility to determine the proton Compton scattering cross section from the reaction $ep\rightarrow ep\gamma$ is studied in the kinematics with a small transversal momentum transfer in the electron leg. 
 With the exception of the region of forward photon directions and extreme photon energies close to the maximum ones or zero, registration of the particles $\gamma$ and $p$ in the final state can enable one to distinguish the sub-process of the Compton scattering from the electron bremsstrahlung background and, owing to complete kinematical reconstruction of each individual event, measure in detail energy and angular dependence of the $\gamma p$-scattering  cross section. 
The count rate expected with a storage ring like NEP having luminosity $L \approx 2 \cdot 10^{35}$ cm$^{-2}\cdot$ sec$^{-1}$ as projected is $\approx 10^3$ events/hour.
Such measurements might help to determine proton polarizabilities with high statistical accuracy.
\end{abstract}

\date{1991}
\maketitle
 
\Large \section{INTRODUCTION}
 
 Photon scattering has served for many years as a very useful tool to study nucleon and nucleus structure at low
and intermediate energies. 
Apart from already common applications for excitation of nucleus levels and giant resonances, recent new possibilities 
are discussed such as exploration free neutron properties or properties of nucleon and nucleon
resonances in the nucleus matter without some distortions caused by initial or final state interactions
~\cite{Christillin, Kohn, Vesper, Arenhovel, Ziegler, Ericson, Schumacher, Schelhaas, Hayward, Austin, Lvov-1}. 
The arrival of a new generation of c.w. electron facilities with energies of a few hundred MeV, corresponding tagged photon beams
with intensity up to $10^8$ s$^{-1}$, and the application of large crystal {$\gamma$-detectors} of high energy resolution allowing for
separation of exclusive final states, puts investigations of $\gamma N$- and $\gamma A$-scattering upon far more solid ground. 
Such investigations are already starting, particularly in Mainz and Saskatoon.

One of the priority problems in the present-day studies of photon scattering by nucleons and nuclei is a more precise measurement 
of cross sections of the $\gamma p$-scattering which is the process lying at the base of understanding and interpreting photon 
scattering by nuclei at energies above giant resonances. 
As an illustration of the hardly satisfactory situation with the present experimental knowledge of $\gamma p$-scattering we can give 
the following example concerning an electric polarizability $\alpha_p$ of the proton which is a parameter determining $\nu$-dependence 
of the $\gamma p$-scattering amplitude and cross  section at low photon energies $\nu$~\cite{Petrunkin, Lvov-2, Lvov-3}: 
\begin{figure}[h!]
\vskip 0.125 in
\includegraphics[trim = 0mm 0mm 0mm 0mm, angle=359.5, width=0.8\linewidth] {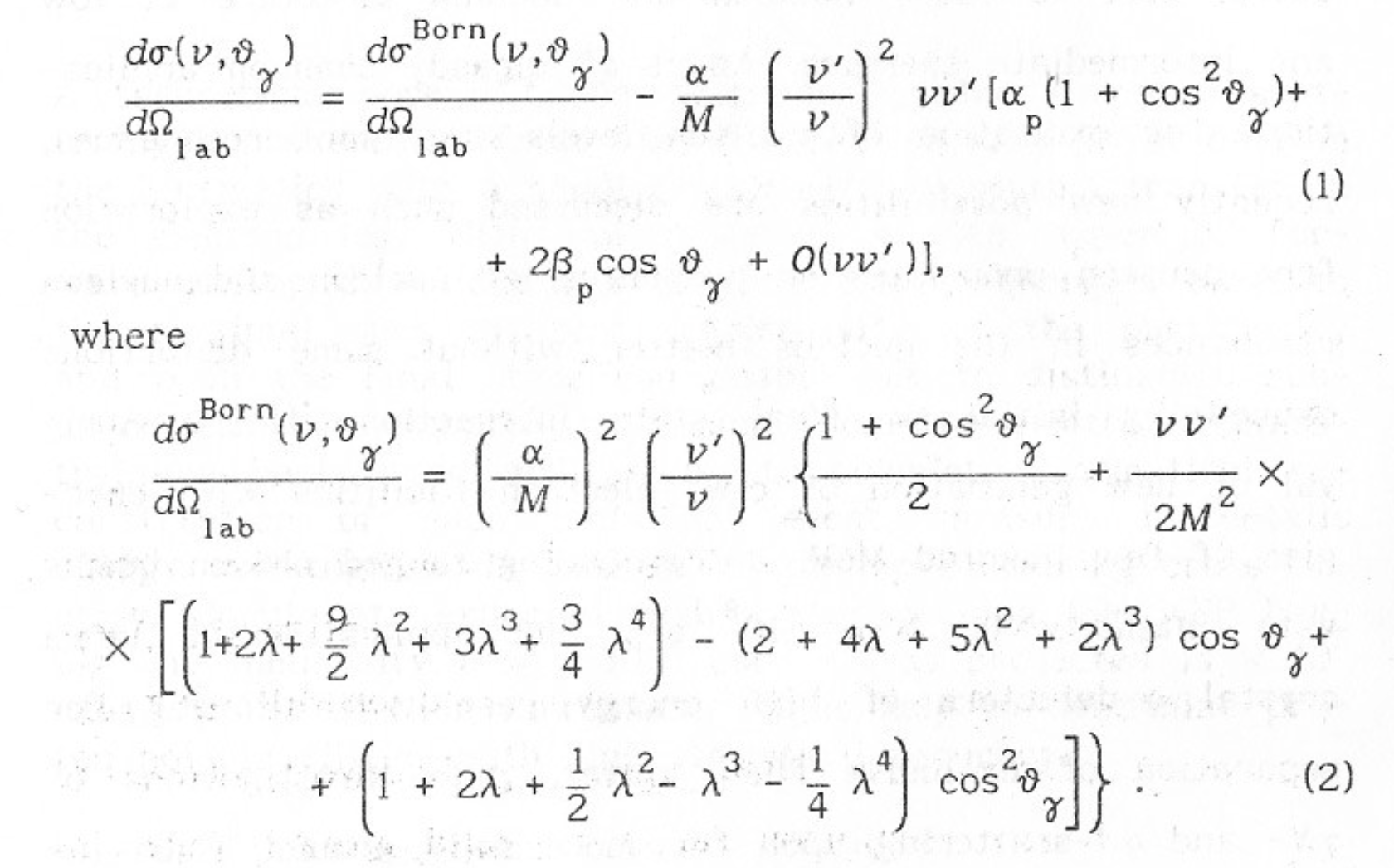}
\end{figure}

Here $M$ and $\lambda$=1.793 are the mass and a.m.m. of the proton, $\alpha$=1/137, $\theta_\gamma$ is the photon scattering angle, 
$\nu$' is the final photon energy, and $\beta_p$ is the proton magnetic polarizability.
\begin{figure}[h!]
\includegraphics[trim = 0mm 0mm 0mm 0mm, angle=0.5, width=0.65\linewidth] {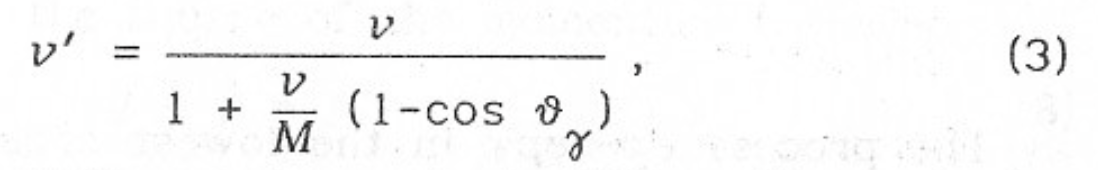}
\end{figure}

The polarizabilities are obtained from the differential cross section of the $\gamma p$-scattering in the region 
$\nu \simeq 50 - 130$ MeV and ranged from $\alpha_p \,=\, (10.7 \pm 1.1) \cdot 10^{-4}$ fm$^3$~\cite{Baranov} 
(old data from Moscow) up to $\alpha_p \approx (15 - 22) \cdot 10^{-4}$ fm$^3$ (very preliminary data 
from Mainz~\cite{Ahrens}) or $\alpha_p \,=\, (17 \pm 3) \cdot 10^{-4}$ fm$^3$ (preliminary data from Illinois~\cite{Nathan}).
Implying that polarizabilities carry very intimate information about valence quarks and pion cloud of the proton~\cite{Petrunkin, Lvov-2}, 
a possibility discussed below to measure with high statistics the p-scattering cross section in the reaction 
$ep \rightarrow ep$ seems to be very attractive. 
This possibility is orientated toward using the internal jet target of the electron storage ring NEP~\cite{Bogdan}.

The low-energy formula (1) predicts that e.g. at $\nu$ = 100 MeV the cross section d$\sigma$/d$\Omega_{lab}$ (90$^o$) 
and the ratio $\sigma (120^o)$ /$\sigma (60^o)$ decrease respectively by 13\% and 17\% when $\alpha_p$
changes from 10 to 15 $ \cdot 10^{-4}$ fm$^3$ (at the fixed sum $\alpha_p + \beta_p \,=\, 14.4 \cdot 10^{-4}$ fm$^3$~\cite{Berg}). 
Hence, obtaining proton polarizabilities is possible from both absolute, and relative measurements
of angular or energy dependence, and the precision $\Delta \alpha \leq 1 \cdot 10^{-4}$ fm$^3$ could be reached by 
measurements of the cross sections within $\leq 2- 3\%$. 
As we will see, the expected luminosity $\simeq 2 \cdot 10^{35}$ cm$^{-2}$s$^{-1}$ of the NEP storage ring is quite 
sufficient to achieve this aim.

\section{KINEMATICS}

The process $ep \rightarrow ep\gamma$ in the lowest order in $\alpha$ is governed by diagrams in Fig.~\ref{fig-1} where notations 
for 4-momenta are also given. 
We discuss its kinematics implying that the initial electron beam of the storage ring has a fixed known energy $E \,=\, p_0$ 
and direction labeled as z-axis, and protons from the jet target are almost at rest. 
Due to conservation of the total 4-momentum,
\begin{figure}[h!]
\includegraphics[trim = 0mm 0mm 0mm 0mm, angle=0, width=0.75\linewidth] {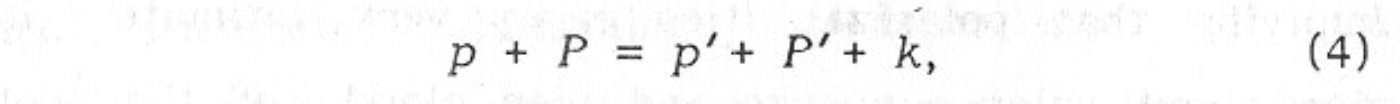}
\end{figure}

and on-mass-shell constraints,
\begin{figure}[h!]
\includegraphics[trim = 0mm 0mm 0mm 0mm, angle=0, width=0.75\linewidth] {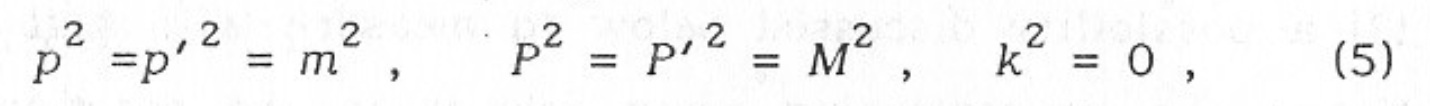}
\end{figure}
where $m$ is the electron mass, among nine values $\vec p$, $\vec k$, $\vec P$' only five are independent. 
Therefore, a registration of the photon and recoil proton with measurements of their energies and angles 
(six values of all) overdetermines the kinematics of the individual event of $ep \rightarrow ep\gamma$. 
Particularly, the relation between the photon energy $\omega  = k_0$ and photon direction $\vec n = \vec k / \omega$,

\begin{figure}[h!]
\includegraphics[trim = 0mm 0mm 0mm 0mm, angle=0, width=0.65\linewidth] {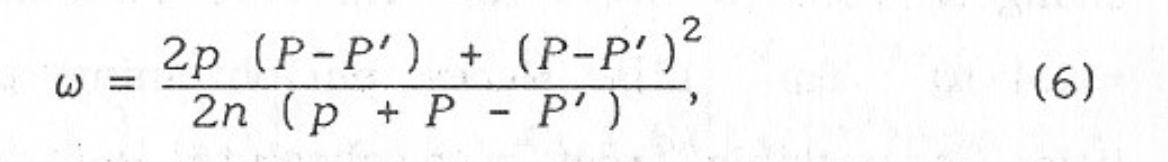}
\end{figure}
where
\begin{figure}[h!]
\includegraphics[trim = 0mm 0mm 0mm 0mm, angle=0, width=0.65\linewidth] {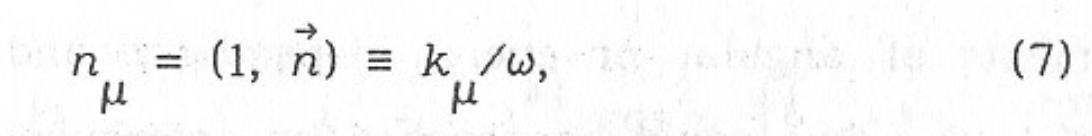}
\end{figure}

allows to refuse oneself from independent measuring of the value $\omega$ or use such measurement 
to define some angles more accurately.

The Compton scattering subprocess interesting for us will dominate when the square of the momentum transfer
\begin{figure}[h!]
\includegraphics[trim = 0mm 0mm 0mm 0mm, angle=0, width=0.8\linewidth] {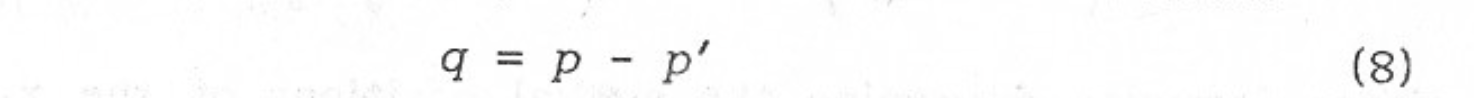}
\end{figure}

is small ($\approx m^2$) or, as is evident from
\begin{figure}[h!]
\includegraphics[trim = 0mm 0mm 0mm 0mm, angle=0.5, width=0.75\linewidth] {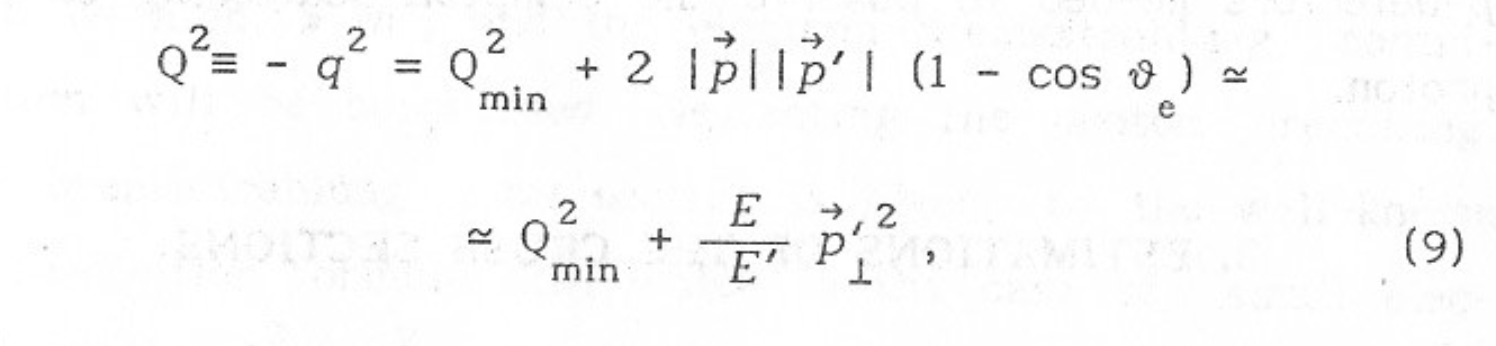}
\end{figure}

when the transversal final electron momentum $\vec p \ '_\perp$ is small. Here
\begin{figure}[h!]
\includegraphics[trim = 0mm 0mm 0mm 0mm, angle=0, width=0.75\linewidth] {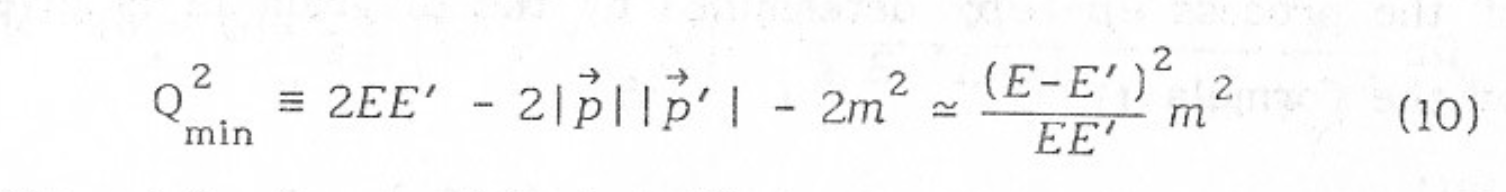}
\end{figure}

is the minimal momentum transfer squared achieved at forward direction of the outgoing electron. 
Under this condition the kinematics for the subprocess of scattering of quasi real photon $\gamma^*$ with the energy
\begin{figure}[h!]
\includegraphics[trim = 0mm 0mm 0mm 0mm, angle=0, width=0.75\linewidth] {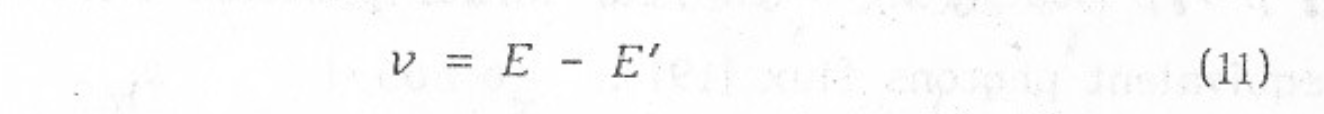}
\end{figure}

is almost the same as in the case of real photons of the energy $\nu$ flying along the z-axis.  
In particular, the energy of the scattered photons $\omega$, kinetic energy of the recoil protons $T_p$ and their polar 
angle $\theta_p$, will be close respectively to

\begin{figure}[h!]
\includegraphics[trim = 0mm 0mm 0mm 0mm, angle=0.25, width=0.75\linewidth] {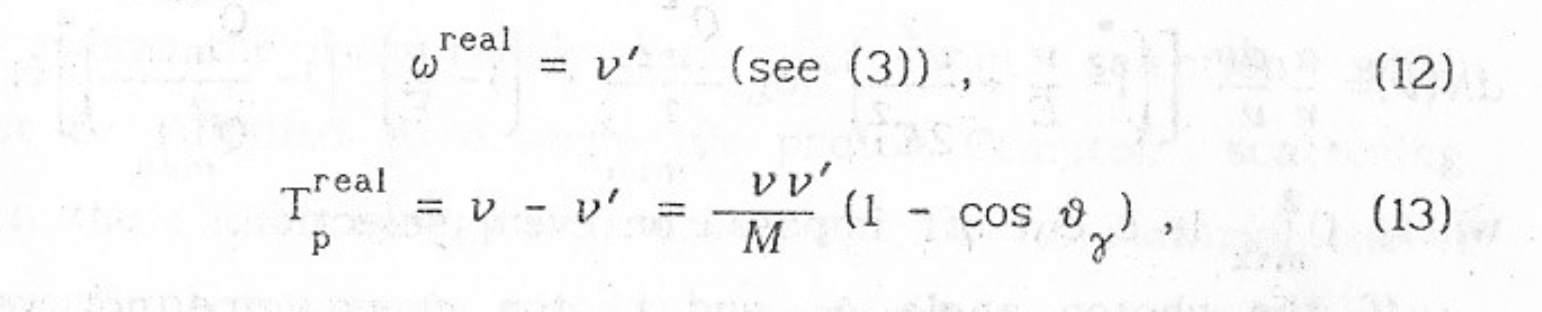}
\end{figure}
\vskip  -0.25 in
\begin{figure}[h!]
\includegraphics[trim = 0mm 0mm 0mm 0mm, angle=0, width=0.65\linewidth] {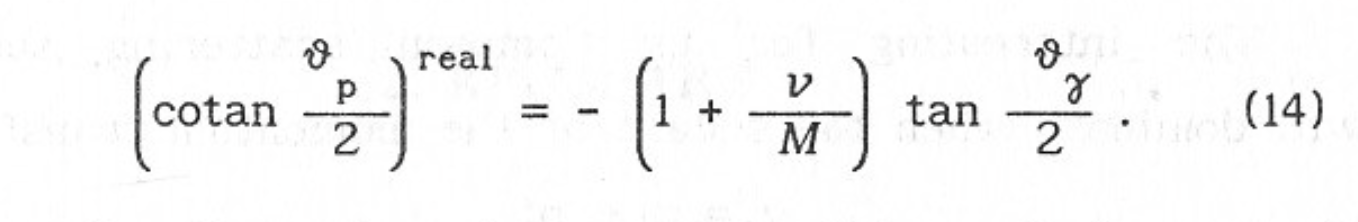}
\end{figure}

\section{ESTIMATIONS OF THE CROSS SECTIONS}

\begin{figure}[!bh]
\includegraphics[width=0.75\columnwidth, trim = 0mm 0mm 0mm 0mm ]{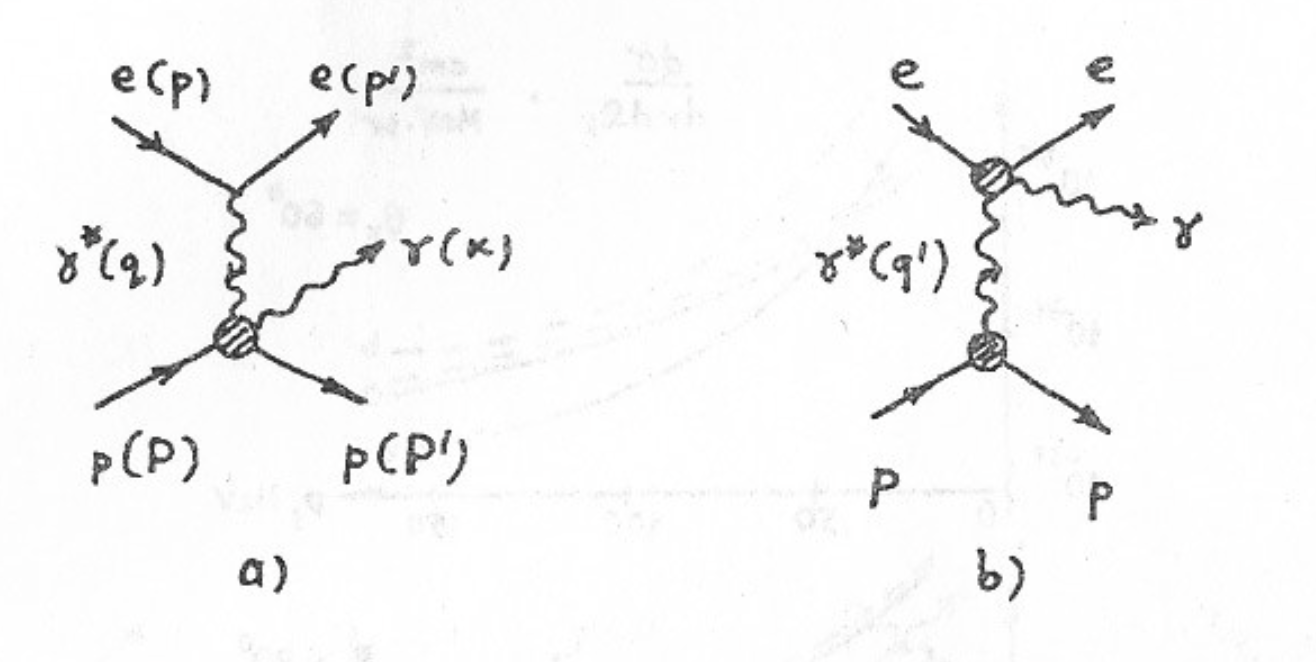} 
\caption{
\large Mechanisms of the reaction $ep \rightarrow ep\gamma$: a) subprocess of photon
scattering by proton; b) electron large angle bremsstrahlung.}
\label{fig-1}
\end{figure}

Under the same condition of small $Q^2$, the cross section of the process $ep \rightarrow ep\gamma$ 
determined by diagram 1a in Fig.~\ref{fig-1} is given by the formula
\begin{figure}[h!]
\includegraphics[trim = 0mm 0mm 0mm 0mm, angle=0, width=0.7\linewidth] {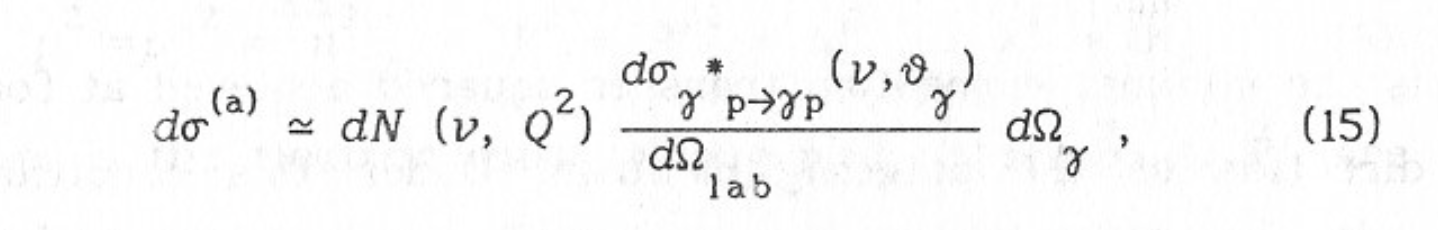}
\end{figure}

and proportional to the differential cross section of the $\gamma^* p \rightarrow \gamma p$ scattering with real initial photons. 
Here $dN$ is the equivalent photons flux~\cite{Budnev} :
\begin{figure}[h!]
\includegraphics[trim = 0mm 0mm 0mm 0mm, angle=0, width=0.8\linewidth] {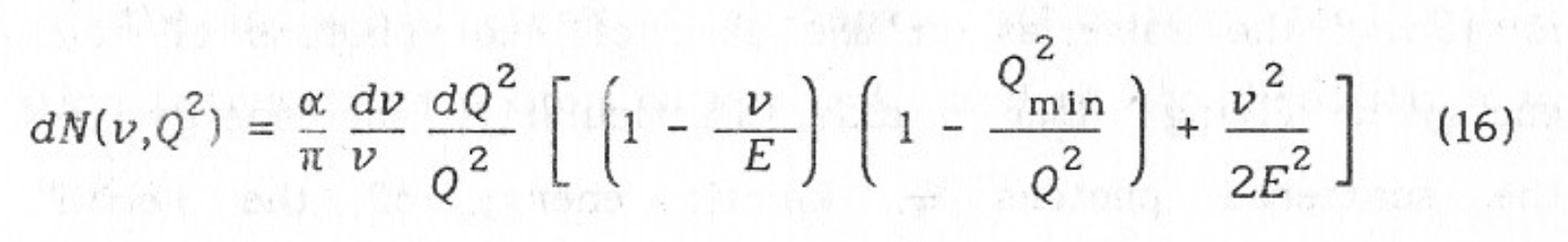}
\end{figure}

or, integrated over $Q^2$,
\begin{figure}[h!]
\includegraphics[trim = 0mm 0mm 0mm 0mm, angle=0, width=0.8\linewidth] {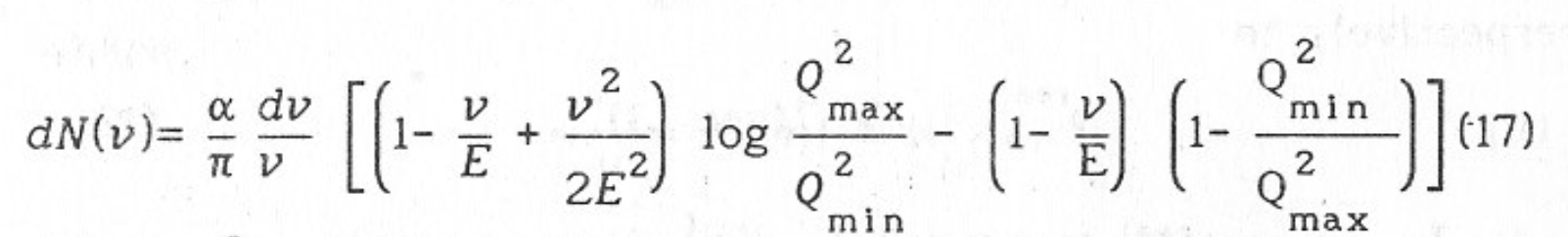}
\end{figure} 
where $Q^2_{max}$ is a cut-off imposed on event selection.

\begin{figure}[!bh]
\includegraphics[width=0.75\columnwidth, trim = {0mm 0mm 0mm 0mm}, clip]{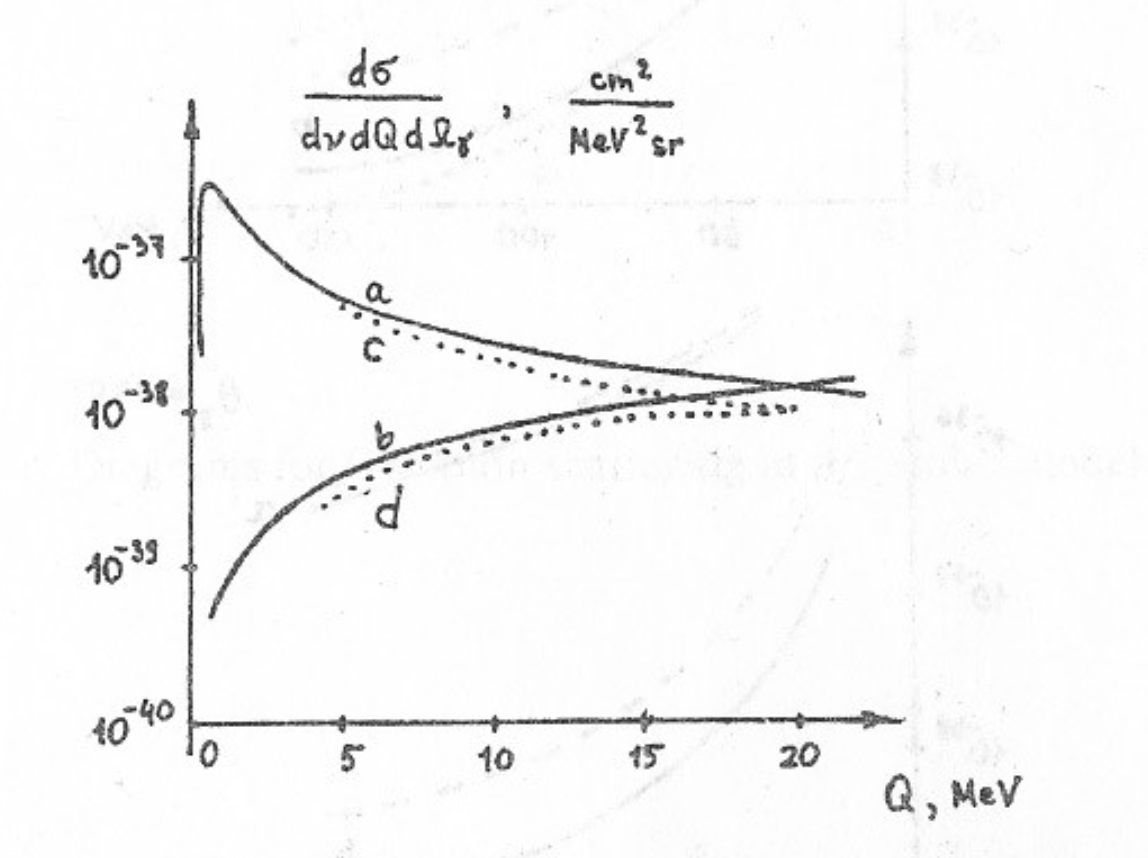} 
\caption{
\large $Q$-dependence of the proton Compton scattering contribution (15)
(curve a) and electron bremsstrahlung (19) (curve b) at $E$= 200 MeV,
$\nu$ = 100 MeV, $\theta_\gamma \,=\, 90^o$. 
Dot curves present the results in the scalar model for photon radiation by electron (curve $c$) and proton (curve $d$).}
\label{fig-2}
\end{figure}

If the photon angle $\theta_\gamma$ and photon energy are not very small, then the 4-momentum transfer squared 
carried out by the virtual photon in diagram 1b in Fig.~\ref{fig-2} and determined by the recoil proton energy
\begin{figure}[h!]
\includegraphics[trim = 0mm 0mm 0mm 0mm, angle=0.75, width=0.55\linewidth] {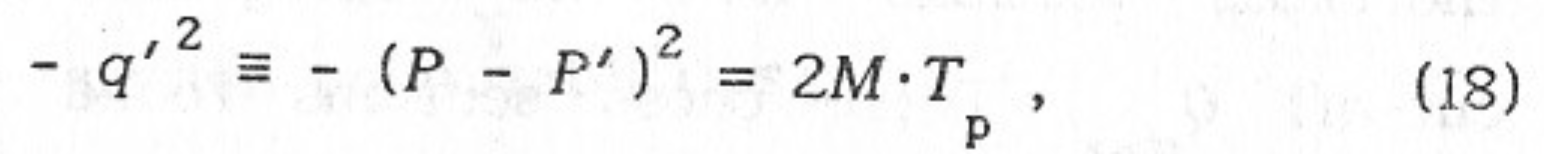}
\end{figure}

will be high, $\gg m^2$, and the electron bremsstrahlung contribution will be suppressed. 
Neglecting the proton recoiling, the bremsstrahlung cross section is given by the well-known 
Bethe-Heitler formula~\cite{Berestetsky},
which in the case of small electron angles $\theta^2_e \ll \theta^2_\gamma$ is reduced to \\

\begin{figure}[h!]
\includegraphics[trim = 0mm 0mm 0mm 0mm, angle=1, width=0.85\linewidth] {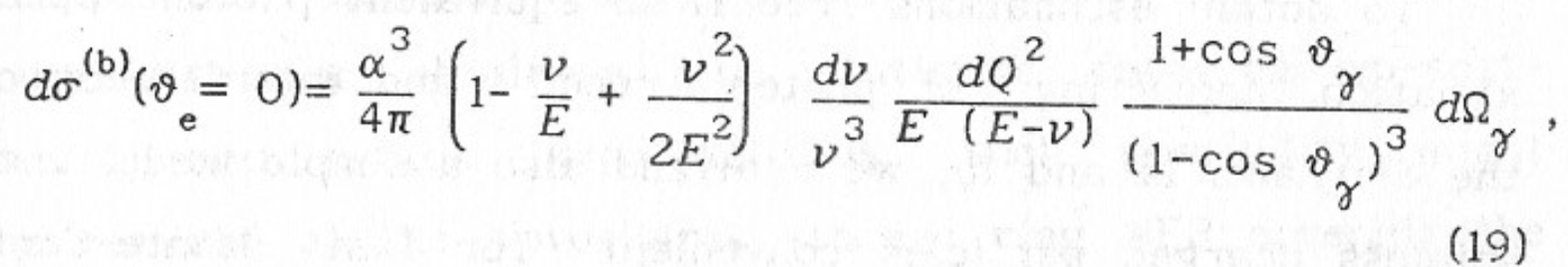}
\end{figure}

so that its ratio to the cross section (15), (1) is equal to
\begin{figure}[h!]
\includegraphics[trim = 0mm 0mm 0mm 0mm, angle=1, width=0.85\linewidth] {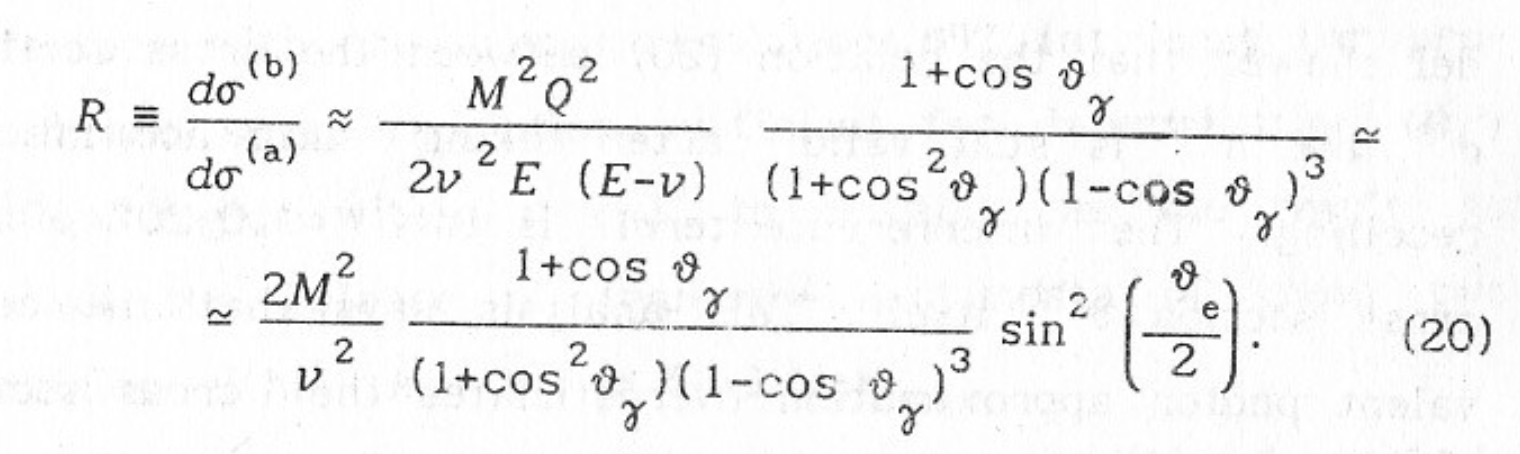}
\end{figure}

At $\nu$ =100 MeV and $\theta_\gamma$ = 90$^o$ this ratio is less than 1 at electron angles up to 
$\theta_e$ = 9$^o$ which corresponds to $Q \sim$ 20 MeV at $E$ = 200 MeV (the NEP energy), see Fig.~\ref{fig-2}. 
The condition $R < 1$ represents the main requirement for event selection which must be fulfilled to 
observe the proton Compton scattering.
When the photon energy $\nu$ is small or close to the maximal one or at small photon angle $\theta_\gamma$ 
this requirement cannot be realized because of the finite resolution of the experimental set-up.

\begin{figure}[!bh]
\includegraphics[width=0.7\columnwidth, trim = {0mm 0mm 0mm 0mm}, clip]{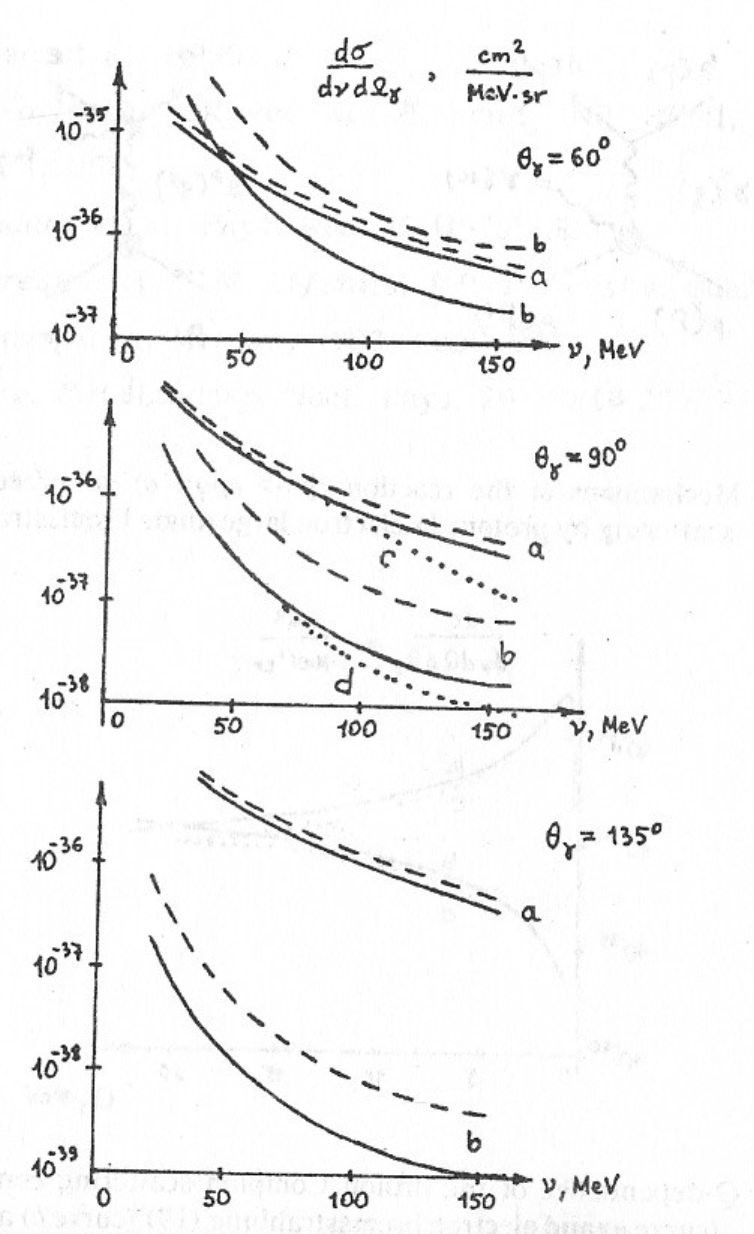} 
\caption{
\large Cross sections for the reaction $ep \rightarrow ep\gamma$ at cut-off $Q_{max}$ = 10 MeV (solid
lines) and $Q_{max}$ = 20 MeV (dashed lines). 
Labels $a$) and $b$) refer to the diagrams contributions (see Fig.~\ref{fig-1}), (15) and (19), respectively, $E$= 200 MeV. 
Dot curves present the cross sections in the scalar model with cut-off $Q_{max}$ = 10 MeV for photon radiation 
by electron (curve $c$) and proton (curve $d$).}
\label{fig-3}
\end{figure}

To estimate count rates it is useful to calculate also the cross sections $d\sigma /d \nu /d\Omega_\gamma$ 
integrated over $Q$ up to some cut-off $Q_{max}$.
These cross sections found with (15), (17), (19) and (21) are depicted in Fig.~\ref{fig-3}.
The typical value of the cross section $\sigma^{(a)}$ is $\approx 10^{-36}$ cm$^2$/MeV/sr so that with the projected 
NEP luminosity $L_e \simeq 2 \cdot 10^{35} \, cm^{-2}s^{-1}$ and $\gamma$-detector aperture 
$\Delta \Omega_\gamma \simeq 0.1$ sr the expected count rate will be $\approx 10^3$ events/hour per 10 MeV 
photon energy interval.
To obtain estimations free from equivalent photon approximation, neglecting of proton recoil and interference of
diagrams la and 1b, we also analyzed a simple model with spinless charged particles convenient for fast Monte-Carlo
simulations (see Appendix). 
The computations within this model showed that the relation (20) between the cross sections 
$\sigma^{(a)}$ and $\sigma^{(b)}$ is still valid after taking into account the recoiling. 
The interference term is only $\simeq$10-20\% of the cross section $\sigma^{(b)}$ itself. 
This analysis says that the equivalent photon approximation overestimates the cross section $\sigma^{(a)}$ by 15-20\%. 
Therefore, a final procedure for obtaining Compton cross sections and proton polarizabilities from the
$ep \rightarrow ep\gamma$ data must be based on more accurate theoretical models than (15). 
Such models can certainly be developed on the base of calculations~\cite{Berg} of the reaction 
$ep \rightarrow ep\gamma$ fulfilled for spin 1/2 particles and model-free low-energy expansion of the 
$\gamma ^* p \rightarrow \gamma p$ subprocess amplitude with virtual photons following from a general 
phenomenological description of polarizable particles~\cite{Lvov-3}. 
The expected accuracy of such models seems to be quite sufficient to allow a precise determination of the 
proton polarizabilities.

\section{CONCLUSION}

The above estimations argue that a study of the reaction $ep \rightarrow ep\gamma$ with the final $\gamma$ 
and $p$-coincidences at internal jet targets of electron storage rings provides an attractive possibility for 
obtaining high statistics data on $\gamma-p$-scattering and proton polarizabilities. 
Such a possibility might be realized at the Novosibirsk storage ring NEP covering the energies up to $\nu  \approx 200$~MeV. 
The realization will require detectors with the energy resolution $\leq 5 \%$ (at least for protons) and space resolution 
sufficient for determination of $\gamma$ and $p$ angles within $\leq 2-3^o$ to be able, as the result, to
control the balance of transversal momenta of these two particles within 
$|\vec k_\perp + \vec P'| \leq 5-10$ MeV.

Comparing this proposal with more traditional methods for studying Compton scattering it should be noted that a
registration and spectrometry of the recoil protons with low energies $\approx 10$~MeV becomes possible 
owing to the very thin target (gas jet). 
Together with the registration of photons it allows one to reconstruct all energies and thereby to have the same 
advantages as with tagged photons. 
At the same time the refusal from registration of final electrons enables one to work with the whole intensity 
of the internal electron beam of the storage ring, thus achieving an effective luminosity in the 
$\gamma^* p$-collisions $L_\gamma \,=\, 10^{33}$ cm$^{-2}$ s$^{-1}$ that is far beyond the possibilities of 
the tagged photon method, and achieving yields $\approx 10^3$~events/hour.
Authors acknowledge M.I. Levchok and M.V. Galynski for pointing out the work~\cite{Berg} and informing us about their own
estimations. 
Thanks also to D.M. Nikolenko, V.G. Raevski and M. Schumacher for useful discussions of the experimental
questions.

\section{APPENDIX. Scalar Model}

\begin{figure}[!bh]
\includegraphics[width=1\columnwidth, trim = {0mm 0mm 0mm 0mm}, clip]{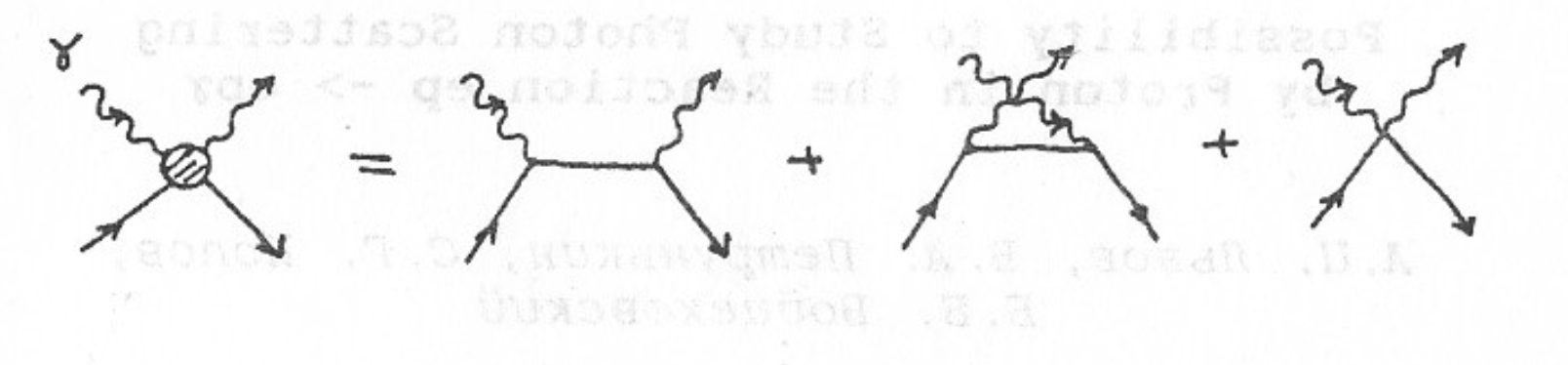} 
\caption{\large Diagrams for Compton scattering in the scalar model.}
\label{fig-4}
\end{figure}

In this model both electrons and protons are considered as spinless point-like particles so that Compton scattering
block is described by diagrams in Fig.~\ref{fig-4}. 
Corresponding $\gamma p$-scattering cross section looks like
\begin{figure}[h!]
\includegraphics[trim = 0mm 0mm 0mm 0mm, angle=1.5, width=0.75\linewidth] {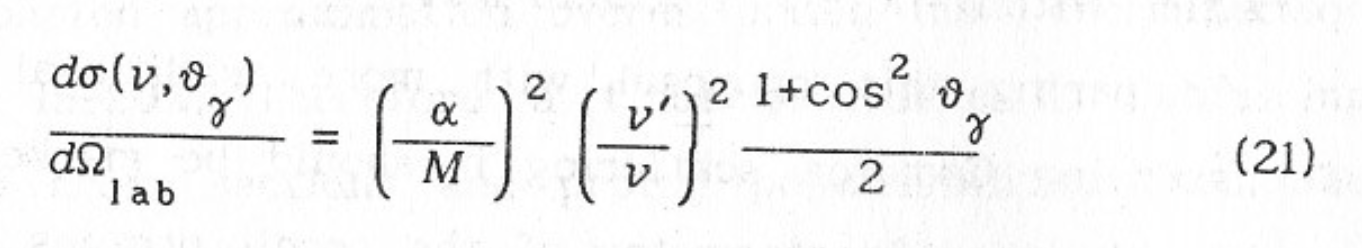}
\end{figure}

and within 10 to 20\% agrees with the real value of the cross section at energies up to 
$\nu \approx 100 - 130$~MeV~\cite{Baranov, Ahrens} where $\Delta$-resonance does not yet dominate.
The scalar model also describes rather well equivalent photons spectrum and electron bremsstrahlung radiation 
off the heavy proton for forward directions of outgoing electrons, with the exception of $\nu$ close to the maximal 
value $E$, yielding instead of (16) and (19) respectively
\begin{figure}[h!]
\includegraphics[trim = 0mm 0mm 0mm 0mm, angle=0.5, width=0.85\linewidth] {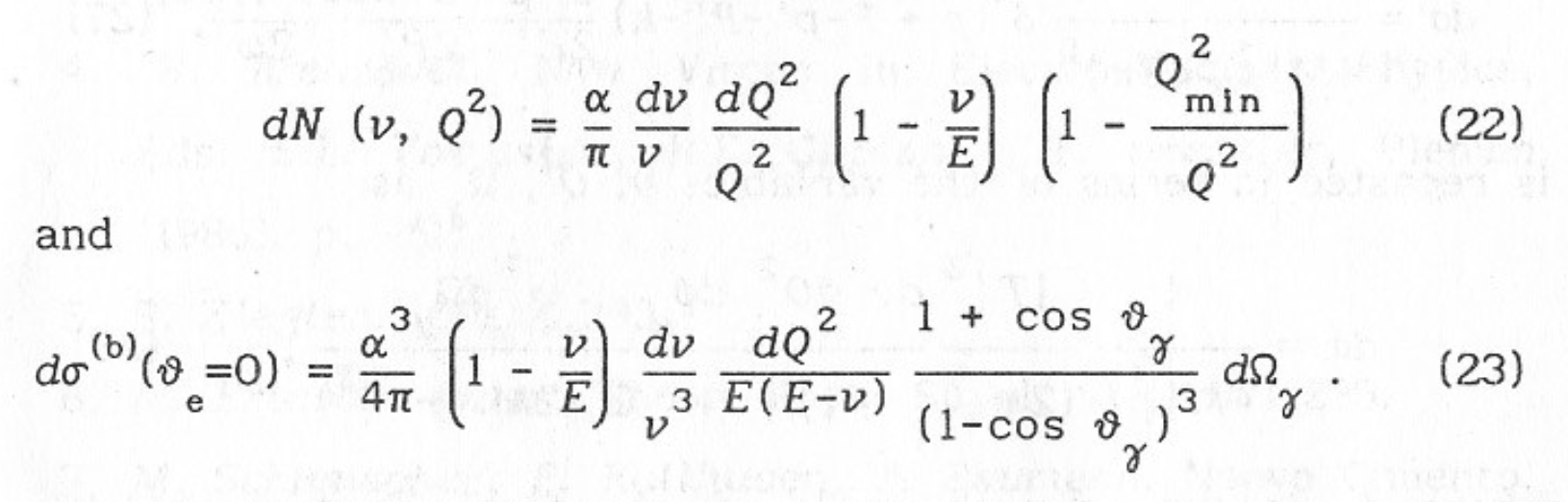}
\end{figure}

Amplitudes of the processes shown in diagrams 1a and 1b, in the frame of the model, are
\vskip 0. in
\begin{figure}[h!]
\includegraphics[trim = 0mm 0mm 0mm 0mm, angle=1, width=0.75\linewidth] {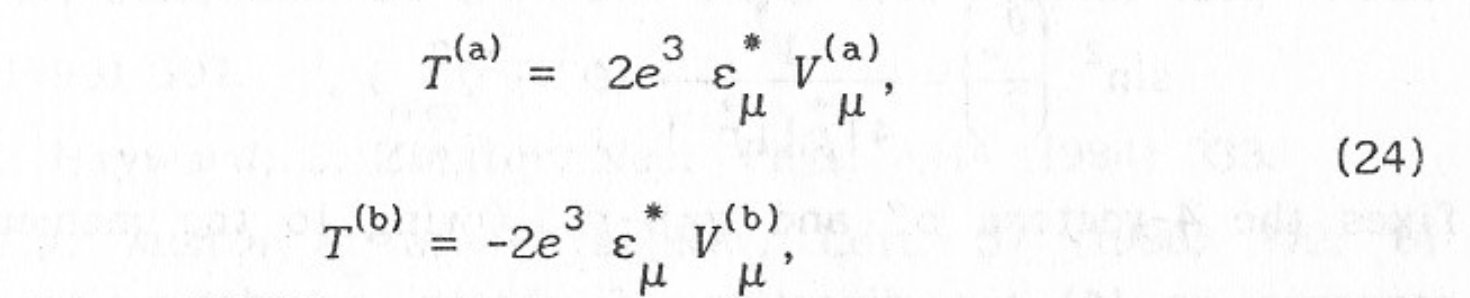}
\end{figure} 

where $e = \sqrt{ 4\pi\alpha}$, $\epsilon_\mu$ is the final photon polarization $(\epsilon k = 0)$ and
\begin{figure}[h!]
\includegraphics[trim = 0mm 0mm 0mm 0mm, angle=1, width=0.85\linewidth] {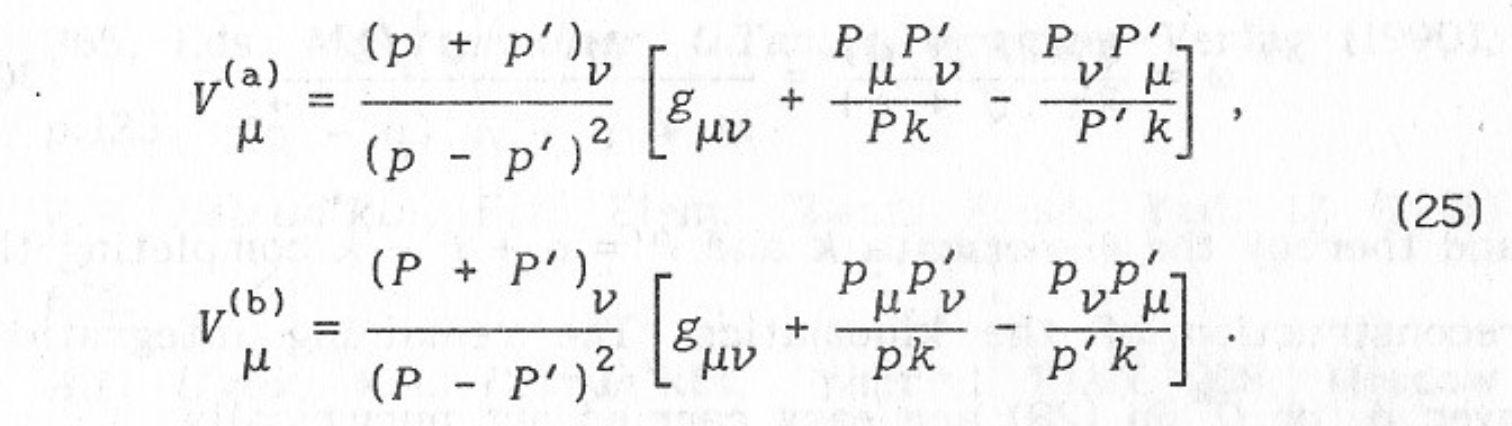}
\end{figure}

Total amplitude squared summarized over photon polarizations,
\begin{figure}[h!]
\includegraphics[trim = 0mm 0mm 0mm 0mm, angle=1, width=0.7\linewidth] {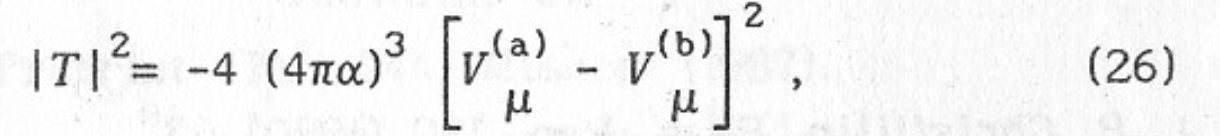}
\end{figure}

is determined by components of the 4-vectors $V^{(a)}$ and $V^{(b)}$ which in turn are easily computed 
from (25) in terms of particles 4-momenta.

A differential cross section of the reaction $ep \rightarrow ep\gamma$,
\begin{figure}[h!]
\includegraphics[trim = 0mm 0mm 0mm 0mm, angle=1, width=0.85\linewidth] {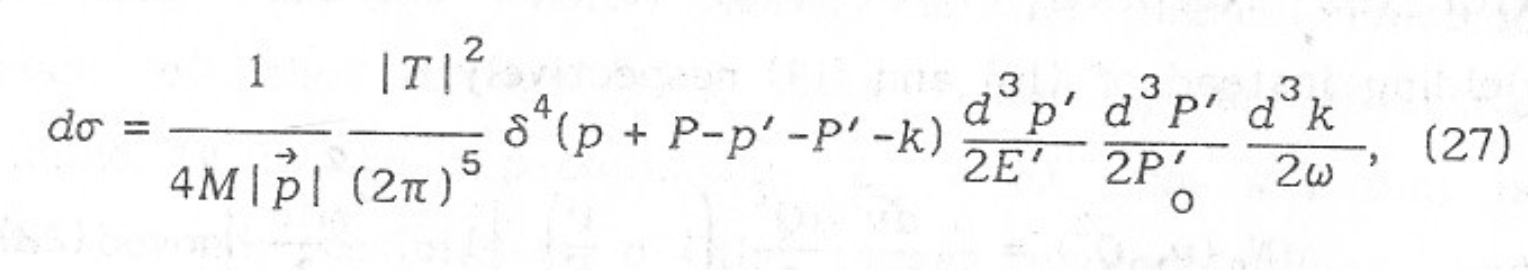}
\end{figure}

is recasted in terms of the variables $\nu$, $Q^2$, $\Omega_\gamma$ as
\begin{figure}[h!]
\includegraphics[trim = 0mm 0mm 0mm 0mm, angle=1, width=0.85\linewidth] {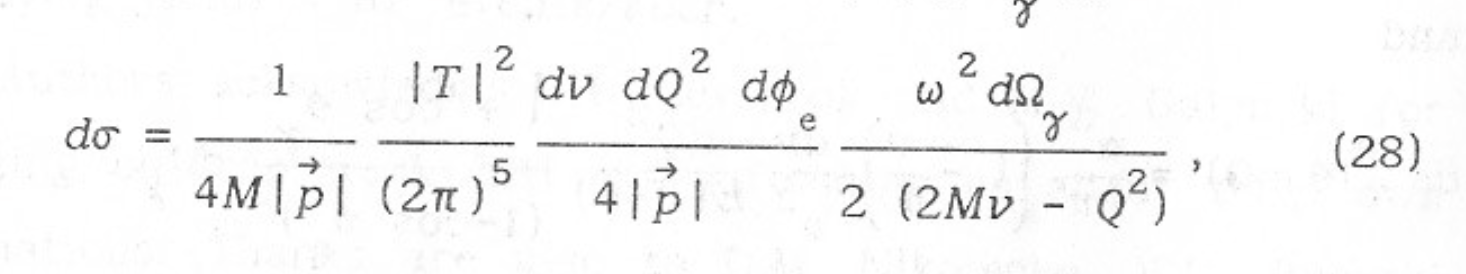}
\end{figure}

where $\phi_e$ is the azimuthal angle of the final electron which, together with the energy 
$E' \,=\, E - \nu$ and the polar angle $\theta_e$,
\begin{figure}[h!]
\includegraphics[trim = 0mm 0mm 0mm 0mm, angle=1, width=0.75\linewidth] {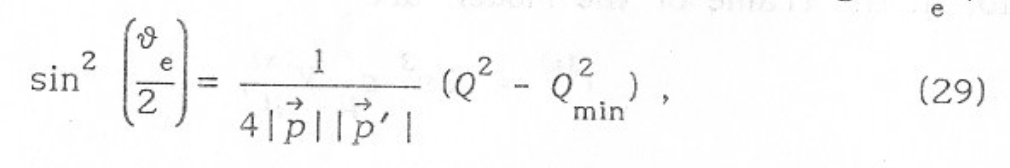}
\end{figure}

fixes the 4-vectors $p '$ and $q \,=\, p-p'$. 
Owing to the momentum conservation (4) the direction of photon $\vec n$ determines the photon energy
\begin{figure}[h!]
\includegraphics[trim = 0mm 0mm 0mm 0mm, angle=1, width=0.75\linewidth] {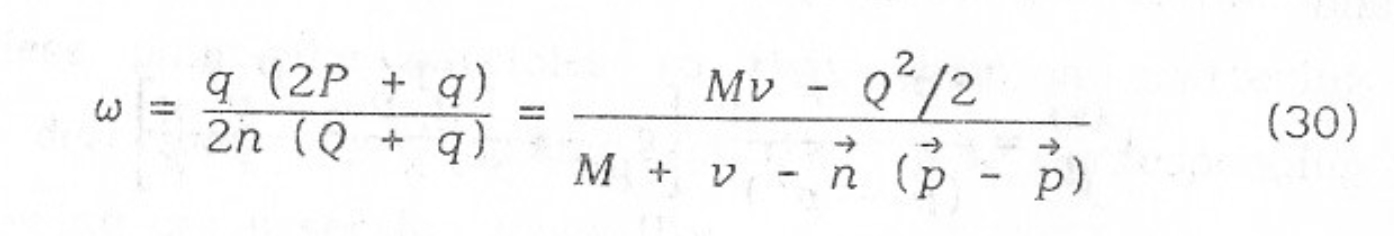}
\end{figure}

and thereby the 4-momenta $k$ and $P' \,=\, q + P - k$ completing the reconstruction of the kinematics. 
The remaining integrations over $\phi_e$ or $Q^2$ in (28) are easily carried out numerically.

\newpage

\end{document}